# Characterization of in-gap states in epitaxial $CoFe_2O_4$(111) layers grown on $Al_2O_3$(111)/Si(111) by resonant inelastic x-ray scattering

Yuki K. Wakabayashi,[1,2,*,a)] Takashi Tokushima,[2,†] Kentaro Kuga,[2] Hiroshi Yomosa,[3] Masaki Oura,[2] Hidenori Fujiwara,[3] Tetsuya Ishikawa,[2] Masaaki Tanaka,[1,4] Takayuki Kiss,[2,3,a)] and Ryosho Nakane[1,5,a)]

[1]*Department of Electrical Engineering and Information Systems, The University of Tokyo, 7-3-1 Hongo, Bunkyo-ku, Tokyo 113-8656, Japan.*
[2]*RIKEN SPring-8 Center, Sayo, Hyogo 679-5148, Japan.*
[3]*Division of Materials Physics, Graduate School of Engineering Science, Osaka University, 1-3 Machikaneyama, Toyonaka, Osaka 560-8531, Japan.*
[4]*Center for Spintronics Research Network, Graduate School of Engineering, The University of Tokyo, 7-3-1 Hongo, Bunkyo-ku, Tokyo 113-8656, Japan.*
[5]*Institute for Innovation in International Engineering Education, The University of Tokyo, 7-3-1 Hongo, Bunkyo-ku, Tokyo 113-8656, Japan.*

[*]Current affiliation: *NTT Basic Research Laboratories, NTT Corporation, 3-1 Morinosato-Wakamiya, Atsugi, Kanagawa 243-0198, Japan*
[†]Current affiliation: *MAX IV Laboratory, Lund University, Fotongatan 2, 224 84 Lund, Sweden*
a)Authors to whom correspondence should be addressed: yuuki.wakabayashi.we@hco.ntt.co.jp, kiss@mp.es.osaka-u.ac.jp, and nakane@cryst.t.u-tokyo.ac.jp

Abstract

We have studied in-gap states in epitaxial $CoFe_2O_4$(111), which potentially acts as a perfect spin filter, grown on a $Al_2O_3$(111)/Si(111) structure by using ellipsometry, Fe $L_{2,3}$-edge x-ray absorption spectroscopy (XAS), and Fe $L_{2,3}$-edge resonant inelastic x-ray scattering (RIXS), and revealed the relation between the in-gap states and chemical defects due to the $Fe^{2+}$ cations at the octahedral sites ($Fe^{2+}$ ($O_h$) cations). The ellipsometry measurements showed the indirect band gap of 1.24 eV for the $CoFe_2O_4$ layer and the Fe $L_{2,3}$-edge XAS confirmed the characteristic photon energy for the preferential excitation of the $Fe^{2+}$ ($O_h$) cations. In the Fe $L_3$-edge RIXS spectra, a band-gap excitation and an excitation whose energy is smaller than the band-gap energy ($E_g$ = 1.24 eV) of $CoF_2O_4$, which we refer to as "below-band-gap excitation (BBGE)" hereafter, were observed. The intensity of the BBGE was strengthened at the preferential excitation energy of the $Fe^{2+}$ ($O_h$) cations. In addition, the intensity of the BBGE was significantly increased when the thickness of the $CoFe_2O_4$ layer was decreased from 11 to 1.4 nm, which coincides with the increase in the site occupancy of the $Fe^{2+}$ ($O_h$) cations with decreasing the thickness. These results indicate that the BBGE comes from the in-gap states of the $Fe^{2+}$ ($O_h$) cations whose density increases near the heterointerface on the bottom $Al_2O_3$ layer. We have demonstrated that RIXS measurements and analyses in combination with ellipsometry and XAS are effective to provide an insight into in-gap states in thin-film oxide heterostructures.

**I. Introduction**

Spin filtering is the spin-dependent electron tunneling through an insulating film having a spin-splitting between the oppositely spin-polarized lower and higher conduction bands. It has generated much attention since it is very useful for the injector and extractor in spintronic devices utilizing the transport of spin-polarized electrons.[1,2] Among insulator films working as spin filters, the ideal inverse spinel ferrite $CoFe_2O_4$ with the Curie temperature $T_C$ of 793 K is promising since its large spin splitting energy (~1 eV) between the lower down-spin and higher up-spin conduction bands potentially causes completely spin-selective electron tunneling at room temperature, namely, the spin-filter effect with 100% efficiency.[3-6] On the other hand, the efficiency of the spin-filter effect was estimated to be less than 4% at room temperature in the past experimental studies on the spin-dependent electron transport through a ferromagnetic multilayer with a thin-film $CoFe_2O_4$ tunnel barrier.[1,2,7,8] These results are thought to be related to imperfection of the $CoFe_2O_4$ tunnel barriers, which leads to the degradation of the magnetic properties and the formation of unwanted in-gap states with opposite polarity to that of the lower down-spin conduction band.[8] However, since electrical measurements cannot clarify the origins, which are related to the electronic structure of $CoFe_2O_4$ tunnel barriers, comprehensive studies on the crystalline structure, electronic structure, and magnetic properties are strongly required.

In $CoFe_2O_4$ ferrite with spinel structure, the crystallographic octahedral ($O_h$) and tetrahedral ($T_d$) sites surrounded by O anions are occupied by Co and Fe cations [Fig. 1(a)]. For the ideal stoichiometry and the ideal $O_h/T_d$ site ratio of 2, the cation distribution is frequently represented by the chemical formula $[Co_{1-y}Fe_y]_{Td}[Fe_{2-y}Co_y]_{Oh}O_4$, where $y$ is the inverse-to-normal ratio called "the inversion parameter" and the valences of Fe and Co cations are assumed to be +3 and +2, respectively. Using this definition, $y = 1$ and 0 represent the inverse and normal spinel structures, respectively, and $0 < y < 1$ represents the coexistence of the inverse and normal spinel structures by antisites between Co and Fe cations in partial regions. From first-principles calculations of the inverse spinel structure ($y = 1$),[5,6] the lower down-spin conduction band is composed of the $3d$ ($t_{2g}$) states of Fe cations at the $O_h$ sites, whereas the higher up-spin conduction band is composed of the $3d$ ($e$) states of Fe cations at the $T_d$ sites, as schematically shown in Fig. 1(b). Hereafter, Fe ($O_h$) and Fe ($T_d$) (Co ($O_h$) and Co ($T_d$)) represent Fe (Co) cations at the $O_h$ and $T_d$ sites, respectively. Theoretically, the electronic band structure with $y = 1$ can develop the spin filter effect with 100% efficiency.[2] However, when $0 < y < 1$, the electronic band structure becomes different from that for the inverse spinel structure,[5,6] and in-gap states can be formed.[7,8] Moreover, even in the inverse spinel structure ($y = 1$), chemical defects arising from the change in the valences of the Fe ($O_h$), and Co ($O_h$) cations, such as $Fe^{2+}$ ($O_h$) and $Co^{3+}$ ($O_h$), will significantly affect the electronic structure.[7] In fact, the magnetization of $CoFe_2O_4$ films, which is related to the electronic structure, was decreased more as the $y$ value decreases from 1 due to the increase in the amounts of $Fe^{2+}$ ($O_h$) and $Co^{3+}$ ($O_h$) chemical defects.[9-11]

Besides the structural and chemical defects described above, experimental studies on inverse spinel ferrite layers, such as $Ni_{1-x}Co_xFe_2O_4$ ($x = 0$-1) and $Fe_3O_4$, formed on different materials have revealed that these layers also contain other structural defects; antiphase boundaries (APBs) and disappearance of $T_d$ sites (a non-ideal $O_h/T_d$ site ratio) near the heterointerface with the bottom material.[9-16] As the amounts of these structural defects increase with decreasing the thickness, the magnetization decreases.[9,14] Thus, these structural defects also lead to the degradation of the magnetic properties and the formation of unwanted in-gap states.

In the previous paper,[9] we quantitatively and systematically characterized the cation distribution and magnetic properties of epitaxial $CoFe_2O_4(111)$ layers with thicknesses $d_{CFO}$ (= 1.4 - 11 nm) grown on a $Al_2O_3(111)/Si(111)$ structure by x-ray absorption spectroscopy (XAS) and x-ray magnetic circular dichroism (XMCD) spectra, aiming at realizing spin injection/extraction into/from Si with an efficient spin filter.[9] Using the configuration-interaction (CI) cluster model, the Fe $L_{2,3}$-edge XAS and XMCD spectra were almost completely reproduced by the weighted sum of the calculated spectra for the $Fe^{3+}$ ($O_h$), $Fe^{3+}$ ($T_d$), and $Fe^{2+}$ ($O_h$) cations. The occupancies of the $Fe^{3+}$ ($T_d$) and $Fe^{2+}$ ($O_h$) cations at $d_{CFO}$ = 11 nm are 38% and 11%, respectively. At $d_{CFO}$ = 1.4 nm, the occupancy of the $Fe^{3+}$ ($T_d$) cations decreases to 27%, and the occupancy of the $Fe^{2+}$ ($O_h$) cations increases to 19%. The $y$ values estimated from these site occupancies are 0.74 at $d_{CFO}$ = 11 nm and 0.54 at $d_{CFO}$ = 1.4 nm. The Co $L_{2,3}$-edge XAS and XMCD spectra were also analyzed in the same manner. The small discrepancy between the experimental and calculated spectra suggested the existence of low-spin-state Co cations and a non-ideal $T_d/O_h$ site ratio. The most important finding was that the $y$ value is strongly related to the magnetic properties; both the magnetization and $y$ value decrease with decreasing $d_{CFO}$, and the magnetic ordering was paramagnetic due to the various complex networks of superexchange interaction at $d_{CFO}$

= 1.4 nm. All the results indicate that when $d_{CFO}$ is thin enough for electron tunneling (~3 nm), the electronic structure significantly changes from the ideal one. However, owing to the measurement principles of XAS and XMCD spectra, it was not clear whether the structural defects and/or chemical defects form the in-gap states in the electronic structure or not. Thus, further studies are needed by using other methods to directly characterize the in-gap states.

$L_{2,3}$-edge resonant inelastic x-ray scattering (RIXS) is a powerful bulk-sensitive technique to study charge excitations, such as band-gap and intra-gap excitations.[17-32] From our previous paper,[9] the $L_{2,3}$ edge XAS spectrum for the $Fe^{2+}$ ($O_h$) cation (chemical defect) is dominant at the photon energy ~708 eV mostly due to an energy shift from the $L_{2,3}$ edge XAS spectra of the $Fe^{3+}$ ($O_h$) and $Fe^{3+}$ ($T_d$) cations. This means that, excitations related to the $Fe^{2+}$ ($O_h$) cations are preferentially strengthened by using the incident photon energy of ~708 eV in the $L_{2,3}$-edge RIXS measurements. In contrast, the $L_{2,3}$ edge XAS spectra of the $Co^{2+}$ ($O_h$), $Co^{2+}$ ($T_d$), and $Co^{3+}$ ($O_h$) cations overlap with each other in the almost entire photon energy range. In addition, considering that the analysis for the Fe cations was more quantitative than that for the Co cations, the in-gap states due to the $Fe^{2+}$ ($O_h$) cations can be simply characterized by using a RIXS spectrum with the preferential excitation of the $Fe^{2+}$ ($O_h$) cations as a reference.

In this paper, we perform Fe $L_{2,3}$-edge RIXS to study the in-gap states induced by the structural and chemical defects in the electronic structure of epitaxial $CoFe_2O_4$ (111) layers with thicknesses $d_{CFO}$ = 1.4 and 11 nm, which are grown on a $Al_2O_3$(111)/Si(111) structure. In preparation to RIXS analysis, we estimate the band gap of the CoFe$_2$O$_4$ layer $d_{CFO}$ = 11 nm by ellipsometry, and measure the Fe $L_{2,3}$ edge XAS spectrum in the CoFe$_2$O$_4$ layer $d_{CFO}$ = 11 nm to confirm the photon energies for the preferential excitations of the Fe$^{3+}$ ($O_h$), Fe$^{3+}$ ($T_d$), and Fe$^{2+}$ ($O_h$) cations. Then, by the analysis with RIXS spectra measured for $d_{CFO}$ = 1.4 and 11 nm, the in-gap states of the Fe$^{2+}$ ($O_h$) cations are clearly characterized by showing incident-photon-energy-dependent peak intensities and an increase in the intensity of the excitation whose energy is smaller than the band-gap energy ($E_g$ = 1.24 eV) of CoF$_2$O$_4$, which we refer to as "below-band-gap excitation (BBGE)" hereafter, with decreasing $d_{CFO}$. From these RIXS measurements and analyses in combination with ellipsometry and XAS, we find that the BBGE comes from the in-gap states of the Fe$^{2+}$ ($O_h$) cations whose density increases near the heterointerface on the bottom Al$_2$O$_3$ layer.

## II. Experiments and Results

### A. Samples

Figure 1(c) shows the sample structure; an epitaxial single-crystalline CoFe$_2$O$_4$(111) layer with a thickness $d_{CFO}$ (= 1.4 or 11 nm) on a 1.4-nm-thick γ-Al$_2$O$_3$(111) buffer layer / 2-nm-thick SiO$_x$ interfacial layer / $n^+$-Si(111) substrate. The samples were grown by pulsed laser deposition, and they are the same as those in our previous paper.[9] Their crystallographic properties of the CoFe$_2$O$_4$(111) layers were characterized by high-resolution transmission electron microscopy (TEM), x-ray diffraction (XRD), atomic force microscopy (AFM), low-energy electron diffraction (LEED), and reflective high-energy electron diffraction (RHEED).[9] The use of the same samples as in the previous study means that a major part of the electronic structure and magnetic properties, such as the correlation between the magnetic properties and occupancies of Fe and Co cations, have been already revealed, which allow us to focus on the in-gap states utilizing those previous results.

**B. Ellipsometry**

Ellipsometry was used to estimate the room-temperature optical band gap of the $CoFe_2O_4$ layer with $d_{CFO}$ = 11 nm. The ellipsometric parameters[33] $\Delta$ and $\psi$, which are related to the Fresnel reflection coefficients,[34] were collected at incident angles of 65°, 70°, and 75° using a rotating compensator ellipsometer (M-2000, JA Woollam),[33] while the photon energy $h\nu$ was swept from 0.738 to 6.36 eV, respectively. Then, the complex refractive index $N\ (=n-\mathrm{i}k)$ spectra were extracted using a least squares regression analysis by fitting the parameters $\Delta$ and $\psi$ of an optical model to the experimental results,[35] where $n$ and $k$ represent the refractive index and extinction coefficient, respectively, and they are functions of $\nu$. To confirm the accuracy of the fittings, experimental data collected at all the incident angles were analyzed. In the fittings, the sample structure was modeled as a 11-nm-thick $CoFe_2O_4$ / 1.4-nm-thick $\gamma$-$Al_2O_3$ / 2-nm-thick $SiO_2$ / Si substrate that was determined by the cross-sectional TEM image,[9] and the $n$ values of $Al_2O_3$, $SiO_2$, and Si were taken from the Woollam database that is originally sourced from the compilation of Palik.[36]

Figure 2(a) shows $N = n - ik$ measured by ellipsometry for the CoFe$_2$O$_4$ layer with $d_{CFO} = 11$ nm, where blue and red curves denote $n$ and $k$, respectively. Using the estimated $k - h\nu$ relation, the indirect band gap of the CoFe$_2$O$_4$ layer was estimated by the Tauc plot.[37,38] For indirect optical transition, the optical absorption depends on the difference between the photon energy and band gap $E_g$ as follows:[37,38]

$$(\alpha h\nu)^{\frac{1}{2}} = A(h\nu - E_g), \qquad (1)$$

where $\alpha = 4\pi k/\lambda$ is the absorption coefficient as a function of $\nu$ and $A$ is a proportionality constant. From Eq. (1), the indirect band gap can be estimated in a $(\alpha h\nu)^{\frac{1}{2}} - h\nu$ plot by extrapolating a fitting line to the zero absorption axis $\alpha = 0$. Figure 2(b) shows $(\alpha h\nu)^{\frac{1}{2}}$ plotted as a function of $h\nu$, where a red solid line is obtained by the $k - h\nu$ relation in (a) and a black dashed line is the fitting in the linear region from 1.24 to 2.00 eV. From the intersection between the black dashed line and $(\alpha h\nu)^{\frac{1}{2}} = 0$ axis, the optical band gap $E_g^{CFO}$ of the CoFe$_2$O$_4$ layer was estimated to be 1.24 eV [Fig. 1(b)], which is consistent with the indirect band gap of CoFe$_2$O$_4$ films estimated by optical measurements (1.18-1.58 eV)[39,40] and electron energy loss spectroscopy (1.3 eV).[41] In CoFe$_2$O$_4$, the band-gap excitations has been assigned to the O 2$p$ → Fe$^{3+}$ ($O_h$) 3$d$ ($t_{2g}$) charge transfer (CT) transition and the Co 3$d$ → Fe$^{3+}$ ($O_h$) 3$d$ ($t_{2g}$) intersite CT-like transition from first principles electronic structure calculations.[5,41]

**C. Fe $L_{2,3}$-edge XAS and RIXS**

XAS and RIXS spectra at room temperature were measured using soft x-rays at the beamline BL17SU in the synchrotron radiation facility SPring-8.[42,43] The energy resolution of monochromator was $E/\Delta E > 10\,000$. The accuracy of the incident photon beam was estimated to be typically ±40 meV. The total instrumental energy resolution for RIXS measurements at a photon energy of 710 eV was estimated to be 0.16 eV from the full width at half maximum of the elastic peak. XAS spectra were measured in the total fluorescence yield (TFY) mode. The incident angle of the photon beam to the sample surface was around 45° and the optical axis of the emission spectrometer was adjusted to perpendicular to the incident beam in the horizontal plane of the incidence.

Figure 3(a) shows a Fe $L_{2,3}$-edge XAS spectrum for the CoFe$_2$O$_4$ layer with $d_{\text{CFO}}$ = 11 nm measured in the TFY mode, where peaked signals in the lower range (707 - 715 eV) and the higher range (720 - 726 eV) correspond to the Fe $L_3$-edge spectrum and $L_2$-edge spectrum, respectively. Both the Fe $L_3$- and $L_2$-edge spectra have large two peak structures originating from the localized 3$d$ state of Fe cations in CoFe$_2$O$_4$,[9,10,44] which are similar to those measured for the same sample in the total electron yield (TEY) mode in our previous study.[9] Thus, despite slight distortion of the XAS spectrum in Fig. 3 due to the self-absorption,[45] our previous analysis of the Fe $L_{2,3}$-edge XAS spectrum with a CI cluster model calculation shown in Fig. 3(b) can be applied; (1) energies marked by $a - f$ in Fig. 3(a) are characteristic energies where each spectrum for each cation (Fe$^{3+}$ ($O_h$), Fe$^{3+}$ ($T_d$), and Fe$^{2+}$ ($O_h$)) has peaks and/or valleys and (2) the signal intensity of the calculated Fe $L_{2,3}$-edge XAS spectrum for the Fe$^{2+}$ ($O_h$) cation becomes maximum and dominates the others at the energy $a$ (= 708.61 eV).[9]

Following (2), excitations related to the $Fe^{2+}$ ($O_h$) cations, in which the electronic states of the $Fe^{2+}$ ($O_h$) cations are initial and/or final states, preferentially appear in a RIXS spectrum at the photon energy *a*.

Figure 4 shows Fe $L_{2,3}$-edge RIXS spectra for the $CoFe_2O_4$ layer with $d_{CFO}$ = 11 nm plotted by the energy-loss scale, where each spectrum was excited at the photon energy *a-f* in Fig. 3. The spectra measured at *a*, *b*, *c*, and *d* are the Fe $L_3$-edge RIXS spectra, whereas those measured at *e* and *f* are the Fe $L_2$-edge RIXS spectra. In all the spectra, the peaked signal at 0 eV is the elastic peak without any energy loss. On the other hand, in the Fe $L_3$-edge RIXS spectra, three peaks indicated by solid lines are inelastic peaks whose energy positions are the same in all the spectra. This confirms that these peaks come from CT or *d-d* transitions. By contrast, in the Fe $L_2$-edge RIXS spectra, three peaks indicated by dotted lines shift by the same amount toward higher photon energy when the excitation photon energy is increased from *e* to *f*.[46] Thus, those three peaks are fluorescence peaks. In this study, the Fe $L_2$-edge RIXS spectra are not analyzed since the fluorescence peaks are not directly related to the band-gap excitation and the BBGE.[47]

Figure 5(a) shows the Fe $L_3$-edge RIXS spectra for the $CoFe_2O_4$ layer with $d_{CFO}$ = 11 nm, where red curves are the magnified experimental spectra in Fig. 4 and blue curves are the sum of black Lorentzian fitting functions with peaks at 0 eV (the elastic peak), 0.5 eV, 1.5 eV, and 3.0 eV (inelastic peaks). Hereafter, each Lorenzian peak is specified by the photon energy at the center value. The most intense inelastic peak at 1.5 eV was possibly due to the CT gap excitation since the peak position is near $E_g^{CFO}$ = 1.24 eV. From the first principles study in ref. 5, the down-spin $Fe^{3+}$ ($O_h$) $3d$ conduction band is composed of the $3d$ ($t_{2g}$) states with the lowest energy and the $3d$ ($e_g$) states with a higher energies by 1-2 eV [Fig. 1(b)]. Following this electronic band structure, the peak at 3.0 eV was assigned to the excitations to the $Fe^{3+}$ ($O_h$) $3d$ ($e_g$) states above the band gap because the energy loss 3.0 eV is larger by 1.5 eV than that (1.5 eV) of the CT gap excitation. In other RIXS measurements for cuprates,[17-21] such CT gap excitation was also observed and thus our assignment is reasonable. On the other hand, it is notable that the peak intensity of the BBGE at 0.5 eV in the spectrum measured at the photon energy $a$ is largest among those measured at the photon energies $a$-$d$. Since the excitations related to the $Fe^{2+}$ ($O_h$) cations are significantly intensified with the photon energy $a$,[9] the BBGE at 0.5 eV probably comes from the in-gap states of the $Fe^{2+}$ ($O_h$) cations. This conclusion is supported by the fact that such excitations related to the in-gap states were observed in carrier doped cuprates[22-24] and manganites.[25]

To verify the assignment of the BBGE due to the in-gap states of the $Fe^{2+}$ ($O_h$) cations, we also measured Fe $L_3$-edge RIXS spectra for the $CoFe_2O_4$ layer with $d_{CFO}$ = 1.4 nm [Fig. 5(b)]. As in the case of $d_{CFO}$ = 11 nm in Fig. 5(a), the experimental spectra are well reproduced by the four fitting functions in all the cases and the full-width at half-maximum of each Lorenzian function at 1.5 or 3.0 eV is almost unchanged at the same excitation photon energy when $d_{CFO}$ is changed from 11 to 1.4 nm. Thus, the electronic structure is basically the same for $d_{CFO}$ = 11 and 1.4 nm. However, the peak intensities of the BBGEs at 0.5 eV in Fig. 5 (b) are drastically larger than those in Fig. 5(a). For the photon energy *a*, when the peak intensity of the BBGE at 0.5 eV is normalized by that at 1.5 eV, the normalized peak intensity increases by a factor of 12.1 when $d_{CFO}$ is decreased from 11 to 1.4 nm. Furthermore, the peak intensity of the BBGE at 0.5 eV is largest in the spectrum measured with the photon energy *a* among those measured with the photon energies *a-d* in Fig. 5(b). As described earlier, the RIXS spectrum for the $Fe^{2+}$ ($O_h$) cations is preferentially excited by the photon energy *a* and the occupancy of the $Fe^{2+}$ ($O_h$) cations at $d_{CFO}$ = 1.4 nm is twice as large as that at $d_{CFO}$ = 11 nm.[9] Therefore, the peak at 0.5 eV comes from the in-gap states of the $Fe^{2+}$ ($O_h$) cations whose density increases near the heterointerface on the bottom $Al_2O_3$ layer.

**III. Discussion**

By the analyses of the experimental results in Section II, it has been revealed that the in-gap states in the $CoFe_2O_4$ layers are formed by the $Fe^{2+}$ ($O_h$) cations (chemical defects) and their density becomes higher near the heterointerface. Although the spin direction of the in-gap states is unknown due to lack of study on the electronic states of the $Fe^{2+}$ ($O_h$) cations in $CoFe_2O_4$, the in-gap states can degrade the spin filter effect. As pointed out in the introduction, this study did not use the Co $L_{2,3}$-edge but the Fe $L_{2,3}$-edge for the RIXS measurements, since the in-gap states of the $Fe^{2+}$ ($O_h$) cations can be simply characterized by using the preferential excitation energy for the $Fe^{2+}$ ($O_h$) cations. Thus, other in-gap states are possibly formed by the structural and chemical defects of Co cations. This can happen since Co cations inherently have a large amount of structural and chemical defects than Fe cations,[9] because Co cations do not have a high selectivity of the $O_h$ sites. Such distribution of Co cations is probably related to the decrease of the $Fe^{3+}$ ($T_d$) cations and the increase of the $Fe^{2+}$ ($O_h$) chemical defects near the heterointerface, which leads to the low inversion parameter $y$ and the significant degradation in the magnetic properties of the thinner $CoFe_2O_4$ layers.

On the other hand, our previous paper[10] showed that the cation distribution and magnetic properties are drastically improved in the $NiFe_2O_4$ layers due to the 100% selectivity of $Ni^{2+}$ cations for the $O_h$ sites; the $y$ value = 0.91 for a $NiFe_2O_4$ layer with $d_{NFO}$ = 3 nm is significantly higher than that (0.74) for the $CoFe_2O_4$ layer with $d_{CFO}$ = 11 nm. In this circumstance, since the in-gap states in the thin $NiFe_2O_4$ layers can be formed only by the structural and chemical defects of Fe cations, their density is expected to be much smaller than that in the $CoFe_2O_4$ layers. As described in the previous paper,[9] the structural defects of Fe cations in the $NiFe_2O_4$ layers mainly originate from disappearance of the $T_d$ sites near the heterointerface, namely, the $O_h/T_d$ site ratio above 2. To decrease the $O_h/T_d$ site ratio toward 2, introduction of tensile strain by a buffer layer is promising.[48] Since the non-ideal $O_h/T_d$ site ratio was found to result in the degradation in magnetic properties, the combination of the XAS and XMCD analyses was very useful to monitor the value of the $O_h/T_d$ site ratio. On the other hand, since the $Fe^{2+}(O_h)$ chemical defects may mostly come from deficiency of O atoms, it should be solved by optimizing the growth conditions of $NiFe_2O_4$ layers. For the optimization to exclude the in-gap states of the $Fe^{2+}(O_h)$ cations, RIXS studies in combination with ellipsometry and XAS are most effective, as demonstrated in this study.

**IV. Conclusion**

We have clearly characterized the in-gap states in the epitaxial CoFe$_2$O$_4$ layers with the different thicknesses ($d_{CFO}$ = 1.4 and 11 nm) by ellipsometry, Fe $L_{2,3}$-edge XAS, and Fe $L_{2,3}$-edge RIXS. The ellipsometry measurements revealed the indirect band gap of 1.24 eV and the Fe $L_{2,3}$-edge XAS spectrum confirmed the characteristic photon energies for the preferential excitations of the Fe$^{3+}$ ($O_h$), Fe$^{3+}$ ($T_d$), and Fe$^{2+}$ ($O_h$) cations. In the Fe $L_3$-edge RIXS spectra, the band-gap excitation and the BBGE were identified by fitting the Lorenzian function to each peak in the spectra and using the indirect band gap value. The preferential excitations of the Fe$^{3+}$ ($O_h$), Fe$^{3+}$ ($T_d$), and Fe$^{2+}$ ($O_h$) cations allowed us to reveal that the BBGE comes from the in-gap states of the Fe$^{2+}$ ($O_h$) cations whose density increases near the heterointerface on the Al$_2$O$_3$ layer. This finding indicates the importance of excluding the Fe$^{2+}$ ($O_h$) cations by interface engineering toward a highly-efficient spin filter through a thin CoFe$_2$O$_4$ tunnel barrier.

By a series of studies through the previous and present papers, we have clarified the detailed material properties of the CoFe$_2$O$_4$ layers by various measurement methods: TEM, XRD, AFM, LEED, RHEED, XAS, XMCD, ellipsometry, and RIXS. Such comprehensive studies are indispensable for engineering various phenomena in insulating magnetic oxides.

**ACKNOWLEDGEMENTS**

This work was partly supported by Grants-in-Aid for Scientific Research (Grants Nos. 26289086), including the Project for Developing Innovation Systems from Ministry of Education, Culture, Sports, Science and Technology (MEXT), the CREST Program of JST (No. JPMJCR1777), and the Spintronics Research Network of Japan (Spin-RNJ). XAS and RIXS experiments were performed at BL17SU in SPring-8 with the approval of RIKEN (proposal no. 20160107). Y. K. W. acknowledges financial support from Japan Society for the Promotion of Science (JSPS) through the Program for Leading Graduate Schools (MERIT) and the JSPS Research Fellowship Program for Young Scientists.

**Figures and Captions**

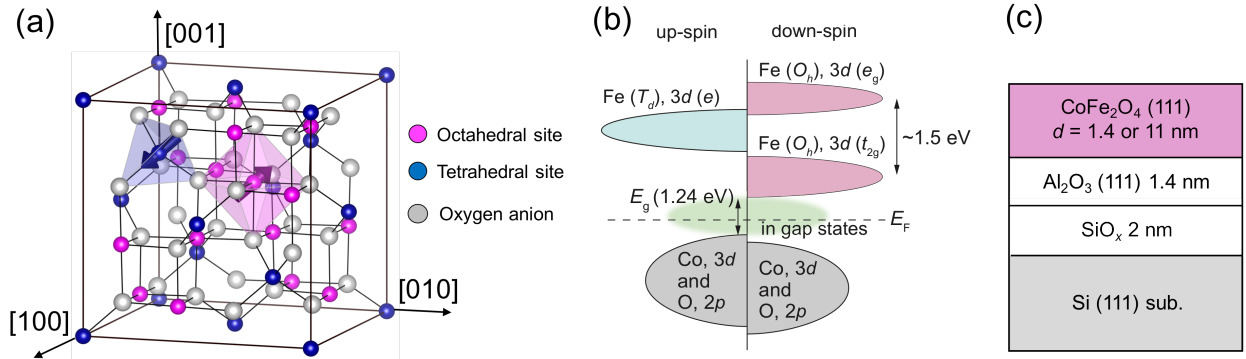

**FIG. 1.** (a) Schematic picture of spinel structure, where small red, small blue, and large gray spheres represent the octahedral ($O_h$) sites, tetrahedral ($T_d$) sites, and oxygen anions, respectively, and blue and red arrows represent the antiferromagnetic coupling between the magnetic moments of cations at the $T_d$ and $O_h$ sites.[9] (b) Schematic up-spin and down-spin density of states (DOS) in inverse spinel $CoFe_2O_4$, where the vertical axis represents the electron energy, $E_F$ represents the Fermi energy, $E_g$ represents the band gap, gray-shaded DOSs formed by Co 3$d$ and O 2$p$ orbitals are located at the up- and down-spin valence-band tops, a red-shaded DOS formed by the Fe($O_h$) 3$d$ ($t_{2g}$) orbital is located at the down-spin conduction-band bottom, a blue-shaded DOS formed by the Fe($T_d$) 3$d$ ($e$) orbital is located at the up-spin conduction-band bottom, and a green-shaded area represents in gap states. (c) Sample structure: (from top to bottom) a $CoFe_2O_4$ layer with the thicknesses $d_{CFO}$ (= 11 or 1.4 nm), a 1.4-nm thick $Al_2O_3$(111) buffer layer, a 2-nm-thick $SiO_x$ inter layer, and a n$^+$-Si(111) substrate.

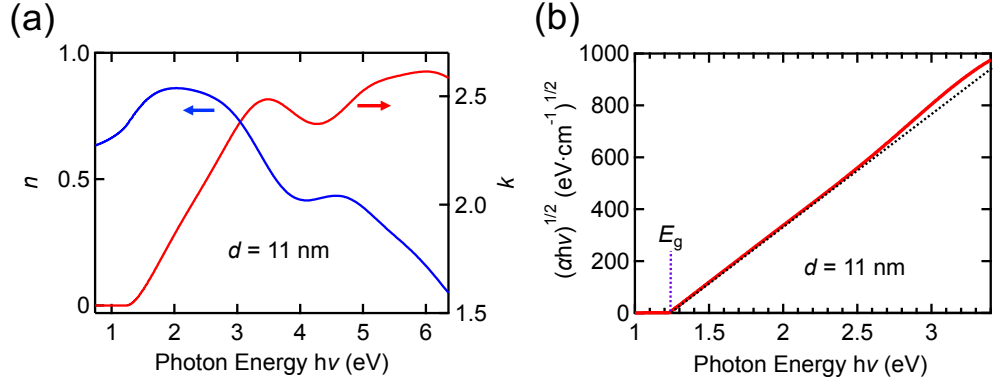

**FIG. 2.** (a) Complex refractive index spectra ($N = n - ik$) for the CoFe$_2$O$_4$ layer with $d_{\text{CFO}}$ = 11 nm measured by room-temperature ellipsometry, where blue and red curves denote $n$ and $k$, respectively. (b) The Tauc plot of the absorption coefficient ($\alpha = 4\pi k/\lambda$) extracted from the extinction coefficient in (a) for an indirect gap, where a black dashed line is the fitting in the linear region from 1.24 to 2.00 eV. In (b), the purple dashed line represents the bang gap.

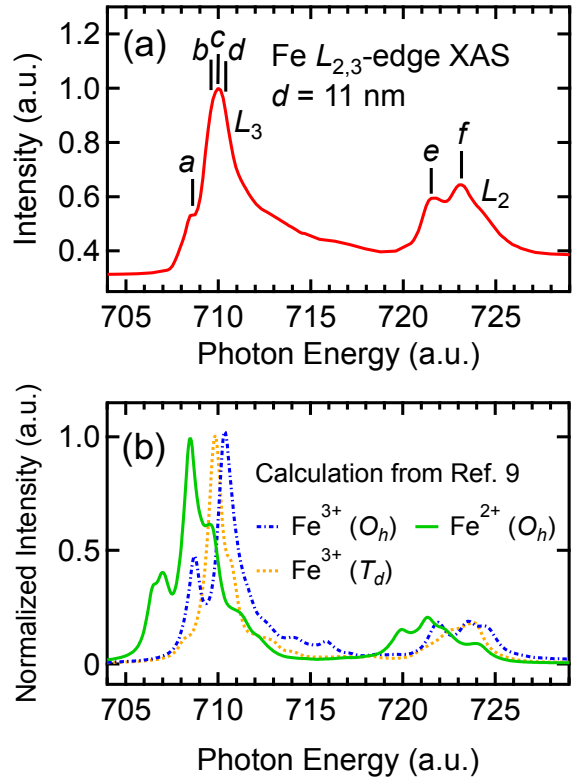

**FIG. 3.** (a) Fe $L_{2,3}$-edge XAS spectrum for the CoFe$_2$O$_4$ layer with $d_{CFO}$ = 11 nm measured at 300 K in the total fluorescence yield (TFY) mode, where $L_3$ and $L_2$ represent peaks originating from the $L_3$-edge XAS spectrum (707 - 715 eV) and the $L_2$-edge XAS spectrum (720 - 726 eV), respectively. Energies marked by $a-f$ are characteristic energies where each spectrum for each cation (Fe$^{3+}$ ($O_h$), Fe$^{3+}$ ($T_d$), and Fe$^{2+}$ ($O_h$)) has peaks and/or valleys. In RIXS measurements, each energy $a$-$f$ is used for the preferential excitation of each cation. (b) Calculated Fe $L_{2,3}$-edge XAS from ref. 9. In (b), each spectrum is normalized by its maximum value, and the dot-dashed, dotted, and solid curves represent the spectra for Fe$^{3+}$ ($O_h$), Fe$^{3+}$ ($T_d$), and Fe$^{2+}$ ($O_h$), respectively.

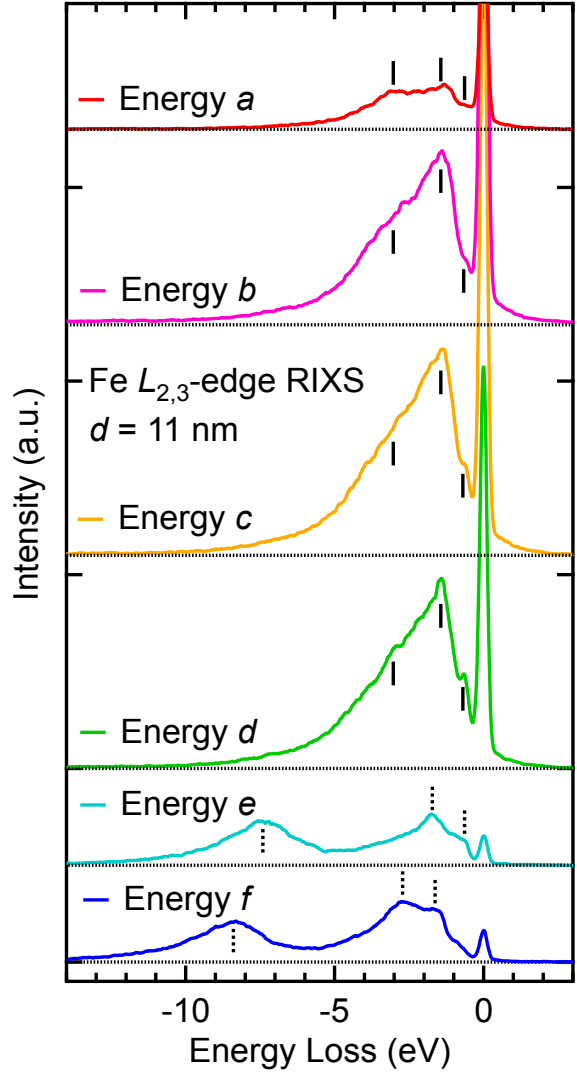

**FIG. 4.** Fe $L_{2,3}$-edge RIXS spectra for the CoFe$_2$O$_4$ layer with $d_{CFO}$ = 11 nm measured at 300 K plotted by the energy-loss scale, where red, pink, orange, green, pale blue, and blue curves are the spectra excited by the photon energies *a*, *b*, *c*, *d*, *e*, and *f* in Fig. 3, respectively. All spectra are normalized by the photon flux. A large peak in each spectrum located at energy loss 0 eV is the elastic peak, solid vertical solid lines in the spectra at the phonon energies *a*, *b*, *c*, and *d* are inelastic peaks from CT or *d-d* transitions, and dashed vertical black lines in the spectra at the phonon energies *e* and *f* are fluorescence inelastic peaks.

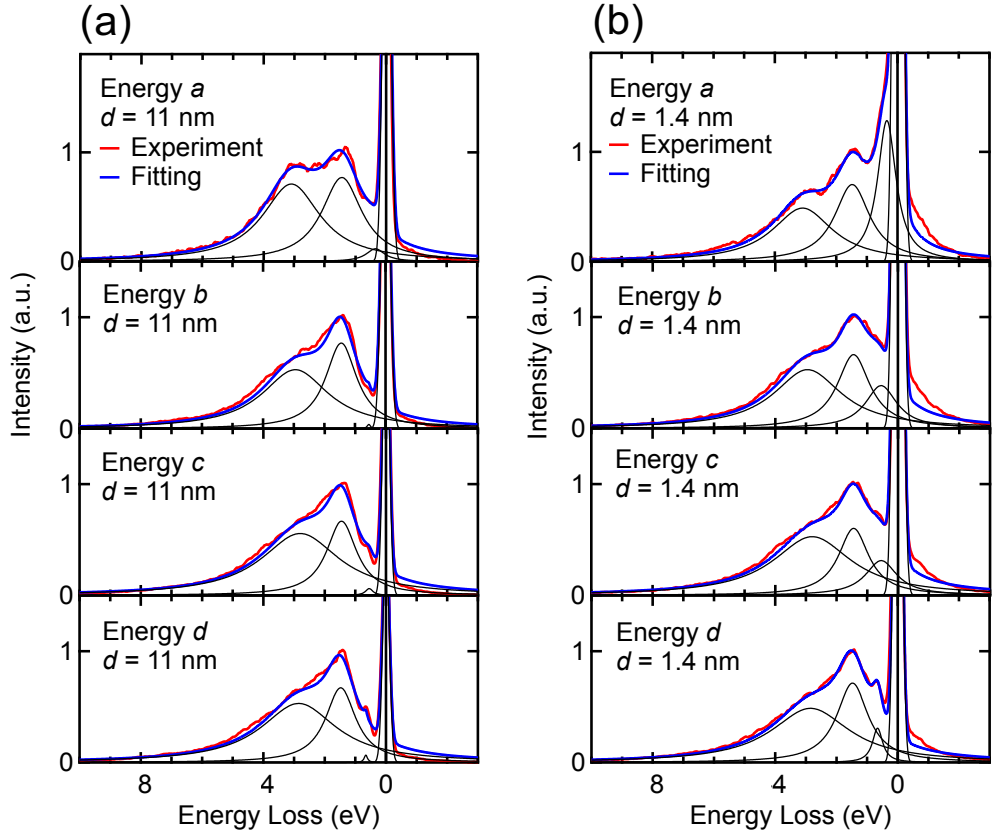

FIG. 5. Fe $L_3$-edge RIXS spectra for the CoFe$_2$O$_4$ layers with $d_{CFO}$ = (a) 11 and (b) 1.4 nm measured at 300 K plotted by the energy-loss scale, where red curves are the experimental spectra excited by the photon energies *a*, *b*, *c*, and *d* in Fig. 3 and blue curves are the sum of black Lorentzian fitting functions with peaks at around 0 eV (the elastic peak), 0.5 eV, 1.5 eV, and 3.0 eV (inelastic peaks). All RIXS spectra are normalized at 1.5 eV for ease of comparison. The red curves in (a) are the magnified experimental spectra in Fig. 4.